\documentclass{article}
\usepackage{amsmath}

\input{tcilatex}

\begin{document}

\author{\textbf{Howard E. Brandt} \\
%EndAName
U.S. Army Research Laboratory, Adelphi, MD\\
hbrandt@arl.army.mil}
\title{\textbf{Conclusive Entangling Probe}}
\maketitle

\begin{abstract}
A design is given for an optimized entangling probe attacking the BB84
(Bennett-Brassard 1984) protocol of quantum key distribution and yielding
maximum information to the probe for a full range of induced error rates.
Probe photon polarization states become optimally entangled with the signal
states on their way between the legitimate transmitter and receiver.
Although standard von-Neumann projective measurements of the probe yield
maximum information on the pre-privacy amplified key, if instead the probe
measurements are performed with a certain positive operator valued measure,
then the measurement results are conclusive, at least some of the time, for
a full range of inconclusive rates.


\end{abstract}

\section{INTRODUCTION}

Recently, a design was presented \cite{PRA-05}, \cite{ModOp-05} for an
optimized entangling probe attacking the BB84 Protocol \cite{Bennett1} of
quantum key distribution (QKD) and yielding maximum information to the
probe. Probe photon polarization states become optimally entangled with the
signal states on their way between the legitimate transmitter and receiver.
Although standard von-Neumann projective measurements of the probe yield
maximum information on the pre-privacy amplified key, it was shown that if
instead the probe measurements are performed with a certain positive
operator valued measure (POVM) \cite{AJP}, \cite{PQE}, \cite{Patent} \cite%
{PRA-1}, \cite{MST} then the measurement results are conclusive, at least
some of the time \cite{JOB-05}. If the inconclusive rate equals the loss
rate of the legitimate receiver (due to attenuation in the key distribution
channel), and only the unambiguous states are relayed by the probe to the
legitimate receiver, then the probe can obtain complete information on the
pre-privacy amplified key, once the bases are announced on the public
channel during reconciliation. The implementation in \cite{PRA-05}, \cite%
{ModOp-05} applied for error rates less than 1/4, thereby allowing
inconclusive rates of the POVM receiver exceeding 1/3. In the present work,
an alternative probe design is presented for which the error rate is less
than 1/3. This enables an inconclusive rate of the POVM receiver ranging
from 0 to 1, matching possible loss rates between the probe and the
legitimate receiver.

In Section 2, the alternative optimized unitary transformation is reviewed,
representing the action of an optimized entangling probe yielding maximum
information on quantum key distribution in the BB84 protocol. In Section 3,
the design is given of the entangling probe for a full range of induced
error rates. In Section 4, analysis is presented for alternatively using the
POVM receiver of \cite{AJP} to measure the probe, thereby unambiguously
discriminating the signal, at least some of the time. Section 5 contains a
summary.

\section{ALTERNATIVE ENTANGLING PROBE}

In the present work, I present an implementation of the optimum unitary
transformation given by Eqs.(158)-(164) of \cite{PRA-05}, however with
restricted parameters such that the corresponding Hilbert space of the probe
reduces from four to two dimensions. In particular, the parameters $\mu $
and $\theta $ are here restricted to%
\begin{equation}
\ \sin \mu =\cos \mu =2^{-1/2},\ \ \ \ \ \cos \theta =1.\ \ 
\end{equation}%
In this case, the entangling probe states $\left\vert \sigma
_{+}\right\rangle $, $\left\vert \sigma _{-}\right\rangle $, $\left\vert
\sigma \right\rangle $, $\left\vert \delta _{+}\right\rangle $, $\left\vert
\delta _{-}\right\rangle $, $\left\vert \delta \right\rangle $, given by
Eqs. (159)-(164) of \cite{PRA-05}, become%
\begin{equation}
\ \left\vert \sigma _{+}\right\rangle =\left\vert \delta _{-}\right\rangle
=4[(1-2E)^{1/2}\left\vert w_{a}\right\rangle -E^{1/2}\left\vert
w_{b}\right\rangle ],\ 
\end{equation}%
\begin{equation}
\ \left\vert \sigma _{-}\right\rangle =\left\vert \delta _{+}\right\rangle
=4[(1-2E)^{1/2}\left\vert w_{a}\right\rangle +E^{1/2}\left\vert
w_{b}\right\rangle ],\ 
\end{equation}%
\begin{equation}
\ \left\vert \sigma \right\rangle =-\left\vert \delta \right\rangle
=4E^{1/2}\left\vert w_{b}\right\rangle ,
\end{equation}%
in which I have made the upper sign choices in Eqs.(159)-(164) of \cite%
{PRA-05}, $E$ is the error rate induced by the probe, and the orthonormal
probe basis vectors $\left\vert w_{a}\right\rangle $ and $\left\vert
w_{b}\right\rangle $ are defined by 
\begin{equation}
\left\vert w_{a}\right\rangle =2^{-1/2}\left( \left\vert w_{0}\right\rangle
+\left\vert w_{3}\right\rangle \right) ,
\end{equation}%
\begin{equation}
\left\vert w_{b}\right\rangle =2^{-1/2}\left( \left\vert w_{1}\right\rangle
-\left\vert w_{2}\right\rangle \right) ,
\end{equation}%
expressed in terms of the orthonormal basis vectors $\left\vert
w_{0}\right\rangle $, $\left\vert w_{3}\right\rangle $, $\left\vert
w_{1}\right\rangle $, and $\left\vert w_{2}\right\rangle $ of \cite{PRA-05}.
Thus, the optimum unitary transformation, Eq.(158) of \cite{PRA-05},
produces in this case the following entanglements for initial probe state $%
\left\vert w\right\rangle $ and incoming BB84 signal photon-polarization
states $\left\vert u\right\rangle $, $\left\vert \overset{\_}{u}%
\right\rangle $, $\left\vert v\right\rangle $, or $\left\vert \overset{\_}{v}%
\right\rangle $, respectively: 
\begin{equation}
\ \left\vert u\right\rangle \otimes \left\vert w\right\rangle
\longrightarrow \frac{1}{4}\left( \left\vert u\right\rangle \otimes
\left\vert \sigma _{+}\right\rangle +\left\vert \overset{\_}{u}\right\rangle
\otimes \left\vert \sigma \right\rangle \right) ,\ \ 
\end{equation}%
\begin{equation}
\ \left\vert \overset{\_}{u}\right\rangle \otimes \left\vert w\right\rangle
\longrightarrow \frac{1}{4}\left( \left\vert u\right\rangle \otimes
\left\vert \sigma \right\rangle +\left\vert \overset{\_}{u}\right\rangle
\otimes \left\vert \sigma _{-}\right\rangle \right) ,\ \ \ 
\end{equation}%
\begin{equation}
\ \left\vert v\right\rangle \otimes \left\vert w\right\rangle
\longrightarrow \frac{1}{4}\left( \left\vert v\right\rangle \otimes
\left\vert \sigma _{-}\right\rangle -\left\vert \overset{\_}{v}\right\rangle
\otimes \left\vert \sigma \right\rangle \right) ,\ 
\end{equation}%
\begin{equation}
\ \left\vert \overset{\_}{v}\right\rangle \otimes \left\vert w\right\rangle
\longrightarrow \frac{1}{4}\left( -\left\vert v\right\rangle \otimes
\left\vert \sigma \right\rangle +\left\vert \overset{\_}{v}\right\rangle
\otimes \left\vert \sigma _{+}\right\rangle \right) .\ 
\end{equation}%
Here, the probe states $\left\vert \sigma _{+}\right\rangle $, $\left\vert
\sigma _{-}\right\rangle $, $\left\vert \sigma \right\rangle $ are given by
Eqs.(2)-(4). The states $\left\vert u\right\rangle $ and $\left\vert \overset%
{\_}{u}\right\rangle $ are orthogonal linearly-polarized photon signal
states in the $\{\left\vert u\right\rangle ,\left\vert \overset{\_}{u}%
\right\rangle \}$ basis, and $\left\vert v\right\rangle $ and $\left\vert 
\overset{\_}{v}\right\rangle $ are orthogonal linearly-polarized photon
signal states in the $\{\left\vert v\right\rangle ,\left\vert \overset{\_}{v}%
\right\rangle \}$ basis, and the two bases are nonorthogonal with $\pi /4$
angle between the linear polarizations of states $\left\vert u\right\rangle $
and $\left\vert v\right\rangle $. In the present case, the maximum
information gain by the probe is again given by 
\begin{equation}
I_{opt}^{R}=\log _{2}\left[ 2-\left( \frac{1-3E}{1-E}\right) ^{2}\right] ,
\end{equation}%
and here $E\leq 1/3$, since $E=1/3$ corresponds to perfect information.

\bigskip

\section{DESIGN OF ENTANGLING PROBE}

Using the same methods presented in \cite{PRA-05}, it can be shown that a
quantum circuit consisting again of a single CNOT gate suffices to produce
the optimum entanglement, Eqs.(7)-(10). Here, the control qubit entering the
control port of the CNOT gate consists of the two signal basis states $%
\{\left\vert e_{0}\right\rangle ,\left\vert e_{1}\right\rangle \}$. In the
two-dimensional Hilbert space of the signal, the basis states $\left\vert
e_{0}\right\rangle $ and $\left\vert e_{1}\right\rangle $, respectively, are
orthogonal and make equal angles of $\pi /8$ with the nonorthogonal signal
states $\left\vert u\right\rangle $ and $\left\vert v\right\rangle $,
respectively. The target qubit entering the target port of the CNOT gate
consists of the two orthonormal linearly-polarized photon polarization basis
states $\left\vert w_{a}\right\rangle $ and $\left\vert w_{b}\right\rangle $
of the probe. When $\left\vert e_{0}\right\rangle $ enters the control port, 
$\{\left\vert w_{a}\right\rangle ,\left\vert w_{b}\right\rangle \}$ becomes $%
\{\left\vert w_{b}\right\rangle ,\left\vert w_{a}\right\rangle \}$, and when 
$\left\vert e_{1}\right\rangle $ enters the control port, $\{\left\vert
w_{a}\right\rangle ,\left\vert w_{b}\right\rangle \}$ remains the same. The
initial unnormalized target state of the probe can, in this case, be shown
to be given by (See Fig. 3 of \cite{PRA-05}):%
\begin{equation}
\ \left\vert A_{2}\right\rangle =(1-2E)^{1/2}\left\vert w_{a}\right\rangle
+(2E)^{1/2}\left\vert w_{b}\right\rangle ,\ 
\end{equation}%
and the unnormalized transition state is given by 
\begin{equation}
\ \left\vert A_{1}\right\rangle =(1-2E)^{1/2}\left\vert w_{a}\right\rangle
-(2E)^{1/2}\left\vert w_{b}\right\rangle .\ 
\end{equation}%
Next, by arguments directly paralleling those of \cite{PRA-05}, using
Eqs.(7)-(10), one has the following correlations between the signal states
and the projected probe states, $\left\vert \sigma _{+}\right\rangle $ and $%
\left\vert \sigma _{-}\right\rangle $:%
\begin{equation}
\ \left\vert u\right\rangle \iff \left\vert \sigma _{+}\right\rangle ,\ \ \
\ \left\vert \overset{\_}{u}\right\rangle \iff \left\vert \sigma
_{-}\right\rangle ,
\end{equation}%
and 
\begin{equation}
\ \left\vert v\right\rangle \iff \left\vert \sigma _{-}\right\rangle ,\ \ \
\ \left\vert \overset{\_}{v}\right\rangle \iff \left\vert \sigma
_{+}\right\rangle .
\end{equation}

The measurement basis for the symmetric von Neumann projective measurement
of the probe must be orthogonal and symmetric about the correlated probe
states, $\left\vert \sigma _{+}\right\rangle $ and $\left\vert \sigma
_{-}\right\rangle $, in the two-dimensional Hilbert space of the probe \cite%
{PRA-05}. Thus, consistent with Eqs.(2) and (3), I define the following
orthonormal measurement basis states: 
\begin{equation}
\ \left\vert w_{+}\right\rangle =2^{-1/2}(\left\vert w_{a}\right\rangle
+\left\vert w_{b}\right\rangle ),
\end{equation}%
\begin{equation}
\ \left\vert w_{-}\right\rangle =2^{-1/2}(\left\vert w_{a}\right\rangle
-\left\vert w_{b}\right\rangle ).
\end{equation}%
Next, one notes that the correlations of the projected probe states $%
\left\vert \sigma _{+}\right\rangle $ and $\left\vert \sigma
_{-}\right\rangle $ with the measurement basis states $\left\vert
w_{+}\right\rangle $ and $\left\vert w_{-}\right\rangle $ are indicated,
according to Eqs.(2), (3), (16), and (17), by the following probabilities: 
\begin{equation}
\ \frac{\ \left\vert \left\langle w_{+}|\sigma _{+}\right\rangle \right\vert
^{2}}{\left\vert \sigma _{+}\right\vert ^{2}}=\ \frac{\ \left\vert
\left\langle w_{-}|\sigma _{-}\right\rangle \right\vert ^{2}}{\left\vert
\sigma _{-}\right\vert ^{2}}=\frac{1}{2}-\frac{E^{1/2}(1-2E)^{1/2}}{(1-E)},
\end{equation}%
\begin{equation}
\ \frac{\ \left\vert \left\langle w_{+}|\sigma _{-}\right\rangle \right\vert
^{2}}{\left\vert \sigma _{-}\right\vert ^{2}}=\ \frac{\ \left\vert
\left\langle w_{-}|\sigma _{+}\right\rangle \right\vert ^{2}}{\left\vert
\sigma _{+}\right\vert ^{2}}=\frac{1}{2}+\frac{E^{1/2}(1-2E)^{1/2}}{(1-E)},
\end{equation}%
consistent with Eqs.(198) and (199) of \cite{PRA-05}, and implying the
following dominant state correlations:%
\begin{equation}
\left\vert \sigma _{+}\right\rangle \iff \left\vert w_{-}\right\rangle ,\ \
\ \left\vert \sigma _{-}\right\rangle \iff \left\vert w_{+}\right\rangle .
\end{equation}%
Next combining the correlations (14), (15), and (20), one thus establishes
the following correlations:%
\begin{equation}
\{\left\vert u\right\rangle ,\left\vert \overset{\_}{v}\right\rangle \}\iff
\left\vert \sigma _{+}\right\rangle \iff \left\vert w_{-}\right\rangle ,
\end{equation}%
\begin{equation}
\ \{\left\vert \overset{\_}{u}\right\rangle ,\left\vert v\right\rangle
\}\iff \left\vert \sigma _{-}\right\rangle \iff \left\vert
w_{+}\right\rangle ,
\end{equation}%
to be implemented by the projective measurement of the probe, as in \cite%
{PRA-05}.

One therefore arrives at the following alternative entangling probe design.
An incident photon coming from the legitimate transmitter is received by the
probe in one of the four signal-photon linear-polarization states $%
\left\vert u\right\rangle $, $\left\vert \overset{\_}{u}\right\rangle $, $%
\left\vert v\right\rangle $, or $\left\vert \overset{\_}{v}\right\rangle $
in the BB84 protocol. The signal photon enters the control port of a CNOT
gate. The initial state of the probe is a photon in linear-polarization
state $\left\vert A_{2}\right\rangle $ entering the target port of the CNOT
gate. The probe photon is produced by a single-photon source and is
appropriately timed with reception of the signal photon by first sampling a
few successive signal pulses to determine the repetition rate of the
transmitter. The photon linear-polarization state $\left\vert
A_{2}\right\rangle $, according to Eq.(12), is given by 
\begin{equation}
\left\vert A_{2}\right\rangle =(1-2E)^{1/2}\left\vert w_{a}\right\rangle
+(2E)^{1/2}\left\vert w_{b}\right\rangle ,\ 
\end{equation}%
and can be simply set for an error rate $E$ by means of a polarizer. In this
way the entangling probe can be tuned to the chosen error rate to be induced
by the probe. The outgoing gated signal photon is relayed on to the
legitimate receiver, and the gated probe photon enters a Wollaston prism,
oriented to separate photon orthogonal-linear-polarization states $%
\left\vert w_{+}\right\rangle $ and $\left\vert w_{-}\right\rangle $, and
the photon is then detected by one of two photodetectors. This is an
ordinary von Neumann projective measurement. If the basis, revealed during
the public basis-reconciliation phase of the BB84 protocol, is $\{\left\vert
u\right\rangle $, $\left\vert \overset{\_}{u}\right\rangle \}$, then the
photodetector located to receive the polarization state $\left\vert
w_{-}\right\rangle $ or $\left\vert w_{+}\right\rangle $, respectively, will
indicate, in accord with the correlations (21) and (22), that a state $%
\left\vert u\right\rangle $ or $\left\vert \overset{\_}{u}\right\rangle $,
respectively, was most likely measured by the legitimate receiver.
Alternatively, if the announced basis is $\{\left\vert v\right\rangle $, $%
\left\vert \overset{\_}{v}\right\rangle \}$, then the photodetector located
to receive the polarization state $\left\vert w_{+}\right\rangle $ or $%
\left\vert w_{-}\right\rangle $, respectively, will indicate, in accord with
the correlations (21) and (22), that a state $\left\vert v\right\rangle $ or 
$\left\vert \overset{\_}{v}\right\rangle $, respectively, was most likely
measured by the legitimate receiver. By comparing the record of probe
photodetector triggering with the sequence of bases revealed during
reconciliation, then the likely sequence of ones and zeroes constituting the
key, prior to privacy amplification, can be assigned. In any case the net
effect is to yield, for a set error rate $E$, the maximum information gain
to the probe, which is given by Eq.(11).

\section{POVM MEASUREMENT OF ENTANGLING PROBE}

Instead of performing a von-Neumann projective measurement of the entangling
probe (using the Wollaston prism along with two photodetectors, as in the
above), one can conclusively detect the two nonorthogonal probe states $%
\left\vert \sigma _{+}\right\rangle $ and $\left\vert \sigma
_{-}\right\rangle $, at least some of the time. For this purpose, the POVM
receiver (See Fig.1 of \cite{AJP}) must simply be set up to distinguish the
nonorthogonal states $\left\vert \sigma _{+}\right\rangle /\left\vert \sigma
_{+}\right\vert $ and $\left\vert \sigma _{-}\right\rangle /\left\vert
\sigma _{-}\right\vert $ (instead of the states $\left\vert u\right\rangle $
and $\left\vert v\right\rangle $ described in \cite{AJP}). For this purpose,
the Wollaston prism in Fig.1 of \cite{AJP} must be aligned to separate the
nonorthogonal states: 
\begin{equation}
\left\vert \overset{\symbol{94}}{e}_{\sigma _{+}+\sigma _{-}}\right\rangle
\equiv \ \frac{\frac{\left\vert \sigma _{+}\right\rangle }{\left\vert \sigma
_{+}\right\vert }\ +\frac{\left\vert \sigma _{-}\right\rangle }{\left\vert
\sigma _{-}\right\vert }}{\left[ \left( \frac{\left\langle \sigma
_{+}\right\vert }{\left\vert \alpha _{+}\right\vert }+\frac{\left\langle
\sigma _{-}\right\vert }{\left\vert \sigma _{-}\right\vert }\right) \left( 
\frac{\left\vert \sigma _{+}\right\rangle }{\left\vert \sigma
_{+}\right\vert }\ +\frac{\left\vert \sigma _{-}\right\rangle }{\left\vert
\sigma _{-}\right\vert }\right) \right] ^{1/2}}
\end{equation}%
and 
\begin{equation}
\left\vert \overset{\symbol{94}}{e}_{\sigma _{+}-\sigma _{-}}\right\rangle
\equiv \ \ \frac{\frac{\left\vert \sigma _{+}\right\rangle }{\left\vert
\sigma _{+}\right\vert }\ -\frac{\left\vert \sigma _{-}\right\rangle }{%
\left\vert \sigma _{-}\right\vert }}{\left[ \left( \frac{\left\langle \sigma
_{+}\right\vert }{\left\vert \alpha _{+}\right\vert }-\frac{\left\langle
\sigma _{-}\right\vert }{\left\vert \sigma _{-}\right\vert }\right) \left( 
\frac{\left\vert \sigma _{+}\right\rangle }{\left\vert \sigma
_{+}\right\vert }\ -\frac{\left\vert \sigma _{-}\right\rangle }{\left\vert
\sigma _{-}\right\vert }\right) \right] ^{1/2}}\ 
\end{equation}%
(instead of $\left\vert \overset{\symbol{94}}{e}_{u+v}\right\rangle $ and $%
\left\vert \overset{\symbol{94}}{e}_{u-v}\right\rangle $, as in \cite{AJP}).
The inconclusive rate $R_{?}$ (or, equivalently, $P_{?}$ in the notation of 
\cite{AJP}) of the POVM receiver is given by \cite{AJP}, \cite{JOB-05} 
\begin{equation}
R_{?}=\ \frac{\left\langle \sigma _{+}|\sigma _{-}\right\rangle }{\left\vert
\sigma _{+}\right\vert \left\vert \sigma _{-}\right\vert }.
\end{equation}%
Next, using Eqs.(2) and (3) in Eq.(26) and solving for $E$, one obtains 
\begin{equation}
E=\ \frac{1-R_{?}}{3-R_{?}}.
\end{equation}%
For this case of measurement of the probe with the POVM receiver, according
to Eq.(27), $E$ can be treated as a parameter ranging from 0 to 1/3 and
determined by a set inconclusive rate $R_{?}$. The conclusive rate $R_{c}$
is given by 
\begin{equation}
R_{c}=1-R_{?}\ .
\end{equation}%
The overlap $Q$ between the states $\left\vert \sigma _{+}\right\rangle $
and $\left\vert \sigma _{-}\right\rangle $ is given by 
\begin{equation}
Q=\ \frac{\left\langle \sigma _{+}|\sigma _{-}\right\rangle }{\left\vert
\sigma _{+}\right\vert \left\vert \sigma _{-}\right\vert },
\end{equation}%
or using Eqs.(26), one obtains 
\begin{equation}
Q=R_{?}.
\end{equation}%
Also, substituting Eq.(27) in Eq.(23), one obtains 
\begin{equation}
\ \left\vert A_{2}\right\rangle =(1+R_{?})^{1/2}\left\vert
w_{a}\right\rangle +2^{1/2}(1-R_{?})^{1/2}\left\vert w_{b}\right\rangle ,\ 
\end{equation}%
in which, since $\left\vert A_{2}\right\rangle $ is not normalized, an
overall factor of $(3-R_{?})^{-1/2}$, appearing in both $\left\vert
A_{2}\right\rangle $ and $\left\vert A_{1}\right\rangle $, is dropped.
Analogously , one obtains 
\begin{equation}
\ \left\vert A_{1}\right\rangle =(1+R_{?})^{1/2}\left\vert
w_{a}\right\rangle -2^{1/2}(1-R_{?})^{1/2}\left\vert w_{b}\right\rangle .\ 
\end{equation}%
According to Eq.(31), the initial state $\left\vert A_{2}\right\rangle $ of
the probe can be tuned to a set inconclusive rate of the POVM receiver. The
reflection coefficient $R_{1}$ of the beamsplitter BS$_{1}$ in the
POVM-receiver in Fig. 1 of \cite{AJP} must, for the case at hand, be given
by 
\begin{equation}
R_{1}=\tan ^{2}\left( \frac{1}{2}\cos ^{-1}Q\right) =\frac{1-Q}{1+Q},
\end{equation}%
or substituting Eq.(30) in Eq.(33), one obtains 
\begin{equation}
R_{1}=\frac{1-R_{?}}{1+R_{?}}.
\end{equation}%
Thus the reflection coefficient $R_{1}$ must be set, according to Eq.(34),
by the set inconclusive rate. This will require a beamsplitter with an
adjustable reflection coefficient.

Finally, it is important to emphasize that, if the photon loss rate, due to
attenuation in the key distribution channel between the probe and the
legitimate receiver, equals the inconclusive rate $R_{?}$, and only the
conclusive states are relayed by the probe to the legitimate receiver, then
the entangling probe together with the POVM receiver can obtain complete
information on the pre-privacy-amplified key, once the polarization bases
are announced in the public channel during reconciliation \cite{JOB-05}.
Also, to counter alteration in the attenuation due to the probe, the
legitimate channel may be replaced by a more transparent one \cite{RMP}. One
may therefore conclude that the BB84 protocol has a vulnerability very
similar to the well-known vulnerability of the B92 (Bennett 1992) protocol 
\cite{JOB-05}, \cite{RMP}. It is also important to emphasize that, because
for the present implementation one has $0\leq E\leq 1/3$, the inconclusive
rate, according to Eq.(27),\ can range here from 0 to 1, and can match a
corresponding loss rate in the channel connecting the probe to the
legitimate receiver. If the inconclusive rate $R_{?}$ is chosen to match the
loss rate in the channel connecting to the legitimate receiver, then the
initial state of the probe must be tuned (using a polarizer located between
the single-photon source and the target entrance port of the CNOT gate) to
the value given by Eq.(31).

\section{CONCLUSION}

The design is determined for an optimized quantum cryptographic entangling
probe attacking the BB84 protocol of QKD and yielding maximum information to
the probe for a full range of induced error rates. Also, it is demonstrated
that if the projective measurement of the probe is replaced by a POVM
receiver, the measurements are conclusive, at least some of the time, for a
full range of inconclusive rates.

\section{ACKNOWLEDGEMENTS}

This work was supported by the U.S. Army Research Laboratory and the Defense
Advanced Research Projects Agency.

\bigskip

\end{document}